\documentclass{ws-procs9x6}
\usepackage{graphics}
\usepackage{graphicx}
\begin{document}

\title{Nuclear density functional constrained by low-energy QCD\footnote{
\uppercase{W}ork supported in part by \uppercase{BMBF} and \uppercase{GSI}}}

\author{Paolo Finelli, Norbert Kaiser, Wolfram Weise}

\address{Physik Department, Technische Universit\"at M\"unchen,\\
D-85747 Garching, Germany}

\author{Dario Vretenar}

\address{Physics Department, University of Zagreb, \\ 
10000 Zagreb, Croatia}  

\maketitle

\abstracts{
We have developed a relativistic point-coupling model of nuclear many-body
dynamics constrained by the low-energy sector of QCD.
The effective Lagrangian is characterized  by density-dependent 
coupling strengths determined by chiral one- and two-pion exchange
(with single and double delta isobar excitations) and by large isoscalar 
background fields that arise through changes of the quark condensate and the
quark density at finite baryon density.
The model has been tested in the analysis of nuclear ground-state 
properties along different isotope chains of medium and heavy nuclei.
The agreement with experimental data is comparable with 
purely phenomenological predictions.
The built-in QCD constraints and the explicit treatment of pion exchange
restrict the freedom in adjusting parameters and functional forms
of density-dependent couplings.
It is shown that chiral pionic fluctuations play an important role
for nuclear binding and saturation mechanism, whereas background fields
of about equal magnitude and opposite sign generate the effective spin-orbit
potential in nuclei.}
 
\section{Introduction}

In this work we would like to investigate the connection between QCD, its
symmetry breaking patterns, and the nuclear many-body problem. 
Usual nuclear structure approaches~\cite{ref1} are 
consistent with the symmetries of QCD (in particular
chiral symmetry) but only functional forms of the interaction terms can be
determined. Model parameters cannot be constrained at the level 
of accuracy required for a quantitative analysis of structure data; they can
only be estimated with the Naive Dimensional Analysis~\cite{nda}.\\
The approach we propose is based on the following ingredients:
\begin{enumerate}
\item The presence of large 
isoscalar background fields which have their origin 
in the in-medium changes of the scalar quark condensate and of the quark
density.
\item The nucleon-nucleon interaction is described by one- and 
two-pion exchange (with medium insertions and delta isobar excitations), 
in combination with Pauli blocking effects.
\end{enumerate} 
The first point has a clear connection with QCD sum rules at finite
density~\cite{ref2}, in which large nucleon self-energies naturally arise in
the presence of a filled Fermi sea of nucleons.\\
The second point is motivated by the observation that, at the nuclear matter
level, the nucleon Fermi momentum $k_f$, the pion mass $m_\pi$ and the 
$\Delta-N$ mass difference represent comparable scales~\cite{ref3}. 
Pionic degrees of freedom are threfore included explicitly
(through density-dependent couplings), in contrast to the phenomenological 
relativistic models, in which effects of iterated one-pion 
and two-pion exchange are treated implicitly through an
{\it effective} scalar field~\cite{ref4}.

\section{The model}

The model is defined by the Lagrangian density\footnote{more
details about density-dependent hadron field theory can be found 
in~\cite{ref5}}
\begin{equation}
   {\mathcal L} = {\mathcal L}_{free} + {\mathcal L}_{4f} 
   + {\mathcal L}_{der} + {\mathcal L}_{coul}
\label{L}
\end{equation}
with the four terms specified as follows:
\begin{eqnarray}
{\mathcal L}_{free} & =  & 
\bar{\psi} [i \gamma_\mu \partial^\mu - M_N] \psi \\
{\mathcal L}_{4f} & =  & 
- \frac{1}{2} G_{S}
   (\hat{\rho})(\bar{\psi}\psi)(\bar{\psi}\psi) 
   -\frac{1}{2} G_{V} (\hat{\rho})(\bar{\psi}\gamma_{\mu}\psi)
   (\bar{\psi}\gamma^{\mu}\psi) \\
~ & ~ & - \frac{1}{2} G_{TS} (\hat{\rho}) 
   (\bar{\psi}\vec{\tau}\psi)\cdot(\bar{\psi}\vec{\tau}\psi) 
   - \frac{1}{2} G_{TV}
   (\hat{\rho})(\bar{\psi}\vec{\tau}\gamma_{\mu}\psi)
   \cdot(\bar{\psi}\vec{\tau}\gamma^{\mu}\psi) \\
{\mathcal L}_{der} & = & 
  -\frac{1}{2} D_{S} (\partial_{\nu}
  \bar{\psi}\psi)(\partial^{\nu}\bar{\psi}\psi) - \frac{1}{2}
  D_{V} (\partial_{\nu}\bar{\psi}\gamma_{\mu}
  \psi)(\partial^{\nu}\bar{\psi}\gamma^{\mu}\psi) \\
 ~ & ~ & -\frac{1}{2} D_{TS} (\partial_{\nu}
  \bar{\psi}\vec{\tau}\psi)\cdot(\partial^{\nu}\bar{\psi}\vec{\tau}\psi) 
  - \frac{1}{2} D_{TV} (\partial_{\nu}\bar{\psi}\vec{\tau}\gamma_{\mu}
  \psi)\cdot(\partial^{\nu}\bar{\psi}\vec{\tau}\gamma^{\mu}\psi) \\
{\mathcal L}_{coul} & = & -eA^{\mu}\bar{\psi}\frac{1+\tau_3}{2}
\gamma_{\mu}\psi -\frac{1}{4} F_{\mu\nu}F^{\mu\nu} \; .
\end{eqnarray}
This Lagrangian is understood to be formally used in the mean-field 
approximation, with fluctuations encoded in density-dependent
couplings $G_i (\hat{\rho})$. Their functional dependence
will be determined from finite-density QCD sum rules
and in-medium chiral perturbation theory 
\begin{equation}
G_i (\hat{\rho}) = G^{(0)} (\hat{\rho}) + G^{(\pi)}  (\hat{\rho}) \; ,
\end{equation}
where the index $i$ labels all isospin-Lorentz structures of Eq.~(\ref{L}). 
The variation of the Lagrangian with 
respect to $\bar{\psi}$, leads to the single-nucleon Dirac equation
\begin{equation}
[\gamma_\mu (i \partial^\mu - V^\mu - V^\mu_R ) - (M + S)]\psi = 0 .
\label{Dirac}
\end{equation}
In addition to the usual self-energies $V^\mu$ and $S$, the density-dependence
of the vertex functions produces the rearrangement contribution\footnote{
the four-velocity $u^\mu$ is defined as 
$(1- {\bf v}^2)^{-1/2}(1,{\bf v})$ (${\bf v} = 0$ in the rest-frame
of the nuclear system)}
\begin{equation}
V_R^{\mu} = 
   u^\mu \left( \frac{1}{2} \frac{\partial G_S}{\partial \hat{\rho}} \rho_s^2 
   +  \frac{1}{2} \frac{\partial G_{TS}}{\partial \hat{\rho}} 
   \vec{\rho}_s \cdot \vec{\rho}_s
   + \frac{1}{2} \frac{\partial G_V}{\partial \hat{\rho}} j^{\nu} j_{\nu}
   + \frac{1}{2} \frac{\partial G_{TV}}
   {\partial \hat{\rho}} \vec{j}^{\nu} \cdot \vec{j}_{\nu} \right) \; .
\end{equation}
The inclusion of this additional term ensures energy-momentum conservation
and thermodynamical consistency. 
For a complete treatment of the density dependent point-coupling
model, the reader is referred to~\cite{ref7}.

\section{Interactions}

\subsection{Self-energies from QCD sum rules}

In leading order, which should be valid below and around saturated 
nuclear matter, the condensate part of the scalar self-energy
\begin{equation}
  \Sigma^{(0)}_S = - \frac{8 \pi^2}{\Lambda_B^2} [
  \langle \bar{q} q \rangle_\rho - \langle \bar{q} q \rangle_0 ]
  = - \frac{8 \pi^2}{\Lambda_B^2}~\frac{\sigma_N}{m_u +m_d} \rho_s \; ,
\end{equation}
is expressed in terms of nucleon sigma-mass term ($\sigma_N =
\langle \bar{N} |m_q \bar{q}q| N \rangle$) and the quark masses.
At the same order, the time-component of the vector self-energy reads
\begin{equation}
   \Sigma_V^{(0)} = \frac{64 \pi^2}{3 \Lambda_B^2} 
  \langle q^\dagger q \rangle_\rho
  = \frac{32 \pi^2}{\Lambda_B^2} \rho \; ,
\end{equation}
where the  quark baryon density is related to that of the nucleons
by $\langle q^\dagger q\rangle_\rho = \frac{3}{2}\rho$.
In both cases, $\Lambda_B \simeq 1~{\rm GeV}$ is the 
characteristic scale (the Borel mass), which approximately separates
the perturbative and non-perturbative energy domains.
\par
For typical values of the nucleon sigma mass term $\sigma_N$ 
($\simeq 50~{\rm MeV}$~\cite{sigma}) and $m_u+m_d$ ($\simeq 12~{\rm MeV}$ at
a renormalization scale of $1$ GeV~\cite{quark}), the in-medium QCD sum rules 
predict large scalar and vector self energies of about equal magnitude
($\simeq 300-400~{\rm MeV}$ in agreement with relativistic 
phenomenological models~\cite{ref1}), and opposite in sign.
Neglecting corrections from higher order condensates, these estimates have 
large uncertainties and the error in the ratio
$\Sigma^{(0)}_S/\Sigma^{(0)}_V \simeq -1$ is about $20\%$.
\par
Given the self-energies arising from the condensate
background, the corresponding equivalent point-coupling strenghts
$G^{(0)}_{S,V}$ are simply determined by
\begin{equation}
   G_S^{(0)} = \frac{\Sigma_S^{(0)}}{\rho_s} \quad {\rm and} \quad
   G_V^{(0)} = \frac{\Sigma_V^{(0)}}{\rho} \; .
\end{equation}  
At leading order the condensate terms of the couplings are constant
and do not contribute to the rearrangement self-energy.

\subsection{Self-energies from in-medium chiral perturbation theory}

In recent years the nuclear matter problem has been extensively studied in the
framework of in-medium chiral perturbation theory. The calculations
have been performed to three-loop order in the energy density and include
one-pion exchange Fock term, one-pion iterated exchange and irreducible
two-pion exchange terms with medium insertions and delta isobar 
excitations~\cite{chiral1}. By adjusting the coupling constants
of few $NN$ contact terms, and encoding short-range
effects not resolved at relevant scales, to the properties of
the empirical saturation point of isospin-symmetric nuclear matter
($\bar{E}_0 = -16~{\rm MeV}$ and $\rho_0= 0.16~{\rm fm}^{-1}$), several aspects
of the problem have been successfully investigated: the
symmetric and asymmetric nuclear matter equation of state (EOS), 
single-particle properties, the low-temperature behaviour, 
and the connection with
non-relativistic nuclear energy density functionals~\cite{chiral1,chiral2}.
\par
\begin{figure}[ht]
    \centering
    \includegraphics*[scale=0.30,angle=0]{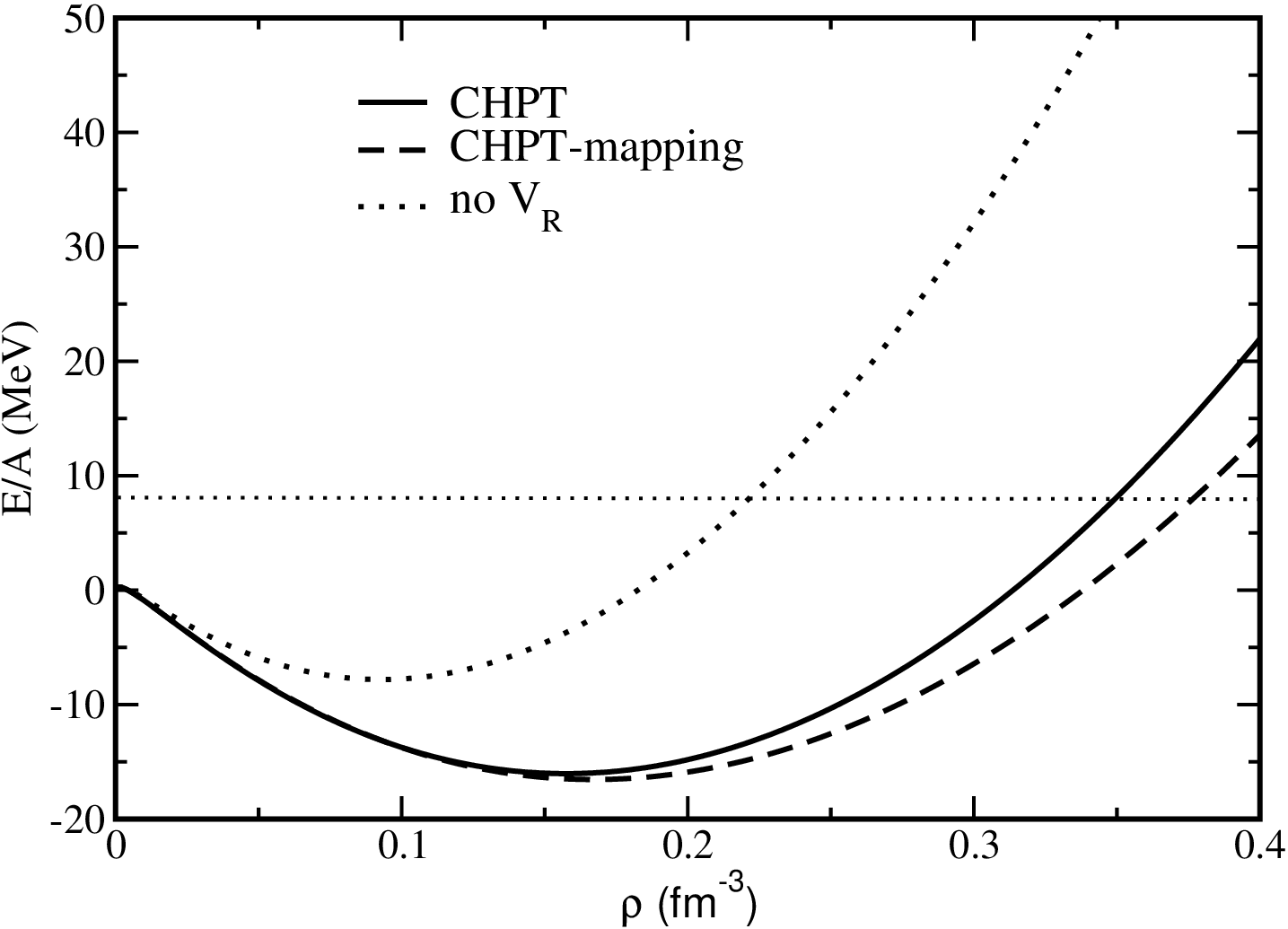}
\caption{Binding energies for symmetric nuclear matter as a function of baryon
density. The solid curve (CHPT) is the EOS calculated by using in-medium
CHPT. The EOS displayed by the dashed curve (CHPT-mapping) is
obtained when the CHPT nucleon self-energies are mapped on the
self-energies of the relativistic point-coupling model with density-dependent
couplings. The dotted curve denotes the latter EOS when rearrangement terms are
neglected. 
\label{fig1}}
\end{figure}

The resulting nucleon self-energies are expressed as expansions in powers of
the Fermi momentum $k_f$. The expansion coefficients are functions 
of $k_f/m_\pi$, the dimensionless ratio of the two relevant small scales:
\begin{equation}
\Sigma^{CHPT} = f(k_f/m_\pi) \frac{k_f^3}{M_N^2} \quad {\rm where} \quad
f(k_f/m_\pi) = \sum_k c_k \left( \frac{k_f}{m_\pi} \right)^k .
\end{equation}  
The density-dependence of the strength parameters is determined by equating
the point-coupling self-energies in the single-nucleon Dirac 
equation~(\ref{Dirac}) with those calculated using in-medium chiral
perturbation theory\footnote{for the definition of
$\rho$, $\rho_s$, $\rho_3$, $\rho_{s3}$ the reader is
referred to~\cite{ref7}}:
\begin{eqnarray}
G_S^{(\pi)} \rho_s             & = & \Sigma_S^{CHPT} (k_f) \\
G_V^{(\pi)} \rho + V^{(\pi)}_R & = & \Sigma_V^{CHPT} (k_f) \\
G_{TS}^{(\pi)} \rho_{s}^3      & = & \Sigma_{TS}^{CHPT} (k_f) \\
G_{TV}^{(\pi)} \rho^3          & = & \Sigma_{TV}^{CHPT} (k_f) .
\end{eqnarray}
In Fig.~(\ref{fig1}) the resulting equation of state 
of isospin-symmetric nuclear matter
is compared with the 
CHPT nuclear matter EOS~\cite{chiral1}. The agreement is 
satisfactory, and the small difference can be attributed 
to the approximations involved (i.e. the momentum dependence
is neglected).

\section{Results}

\begin{figure}[ht]
    \centering
    \includegraphics*[scale=0.43,angle=0]{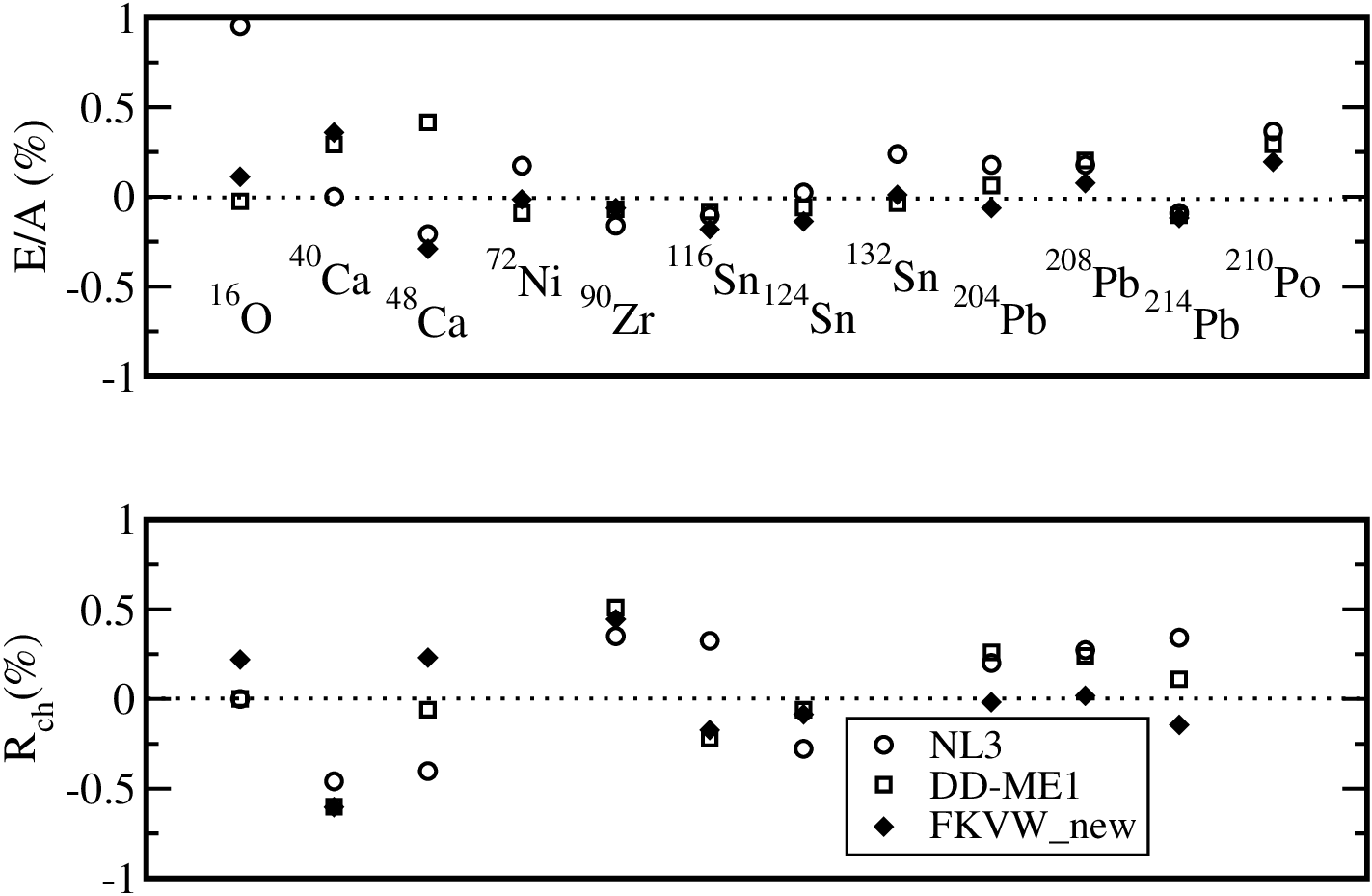}
\caption{Relative errors for the observables
binding energies (upper panel), and charge radii (lower panel), in comparison
with the results of NL3 and DD-ME1. \label{fig2}}
\end{figure}

\begin{figure}[ht]
    \centering
    \includegraphics*[scale=0.30,angle=0]{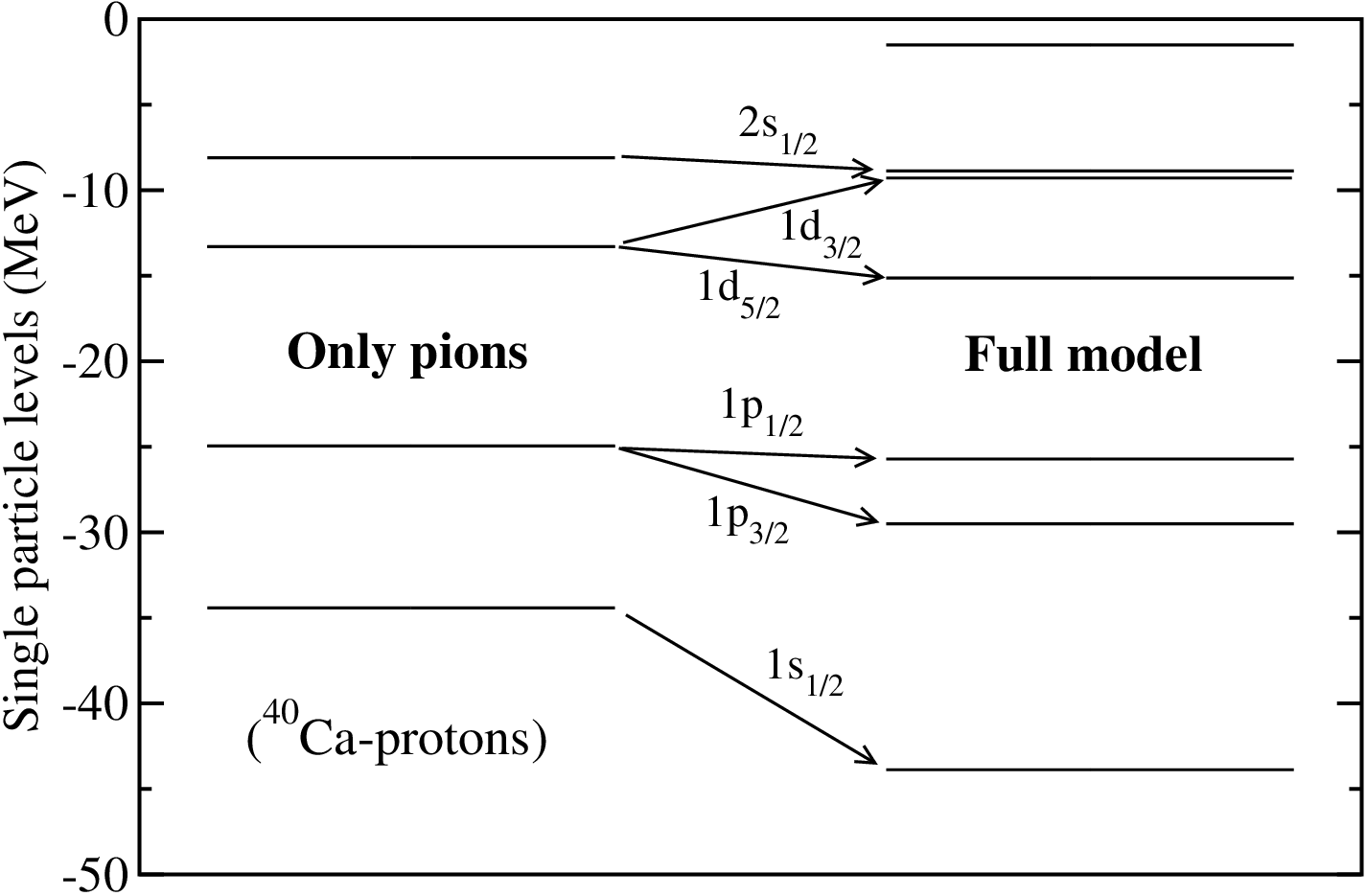}
\caption{Single proton levels for $^{40}$Ca with only pionic fluctuations
(left panel) and with the inclusion of background fields (right panel).
In the latter case the degeneracy of the p- and d-levels is removed.   
\label{fig3}}
\end{figure}

In this section the relativistic point-coupling model will be applied
to the description of ground-state properties of finite nuclei. 
The density-dependent coupling strenghts will be determined
in a {\it least-squares fit} to observables ($\mathcal{O}$) 
of a rather large set of nuclei along the valley of $\beta$-stability 
(from $^{16}$O to $^{214}$Pb):
\begin{equation}
\chi^2 = \sum\limits_{nuclei} \left( \sum\limits_{observ.} \left|
\frac{\langle \mathcal{O}_{th} - \mathcal{O}_{exp} \rangle}{weights} \right|^2
 \right).
\end{equation}
In Fig.~\ref{fig2} the overall agreement is shown in comparison 
with two standard relativistic meson-exchange parametrization: 
NL3~\cite{NL3}, with
explicit higher order self-interaction terms
for the $\sigma$-meson, and DD-ME1~\cite{DD}, with phenomenological 
density-dependent
couplings strenghts.
The interaction determined by the {\it least-squares fit} procedure
contains pionic fluctuations on top of the large background fields.
Nonetheless, it is interesting to analyze the role of pionic fluctuations,
without the presence of background fields, and how these fields
modify the single-particle spectra in finite systems.
In Fig.~\ref{fig3} we display the single proton levels
in $^{40}$Ca without and with background fields. 
It is obvious that the the spin-orbit potential 
\begin{equation}
V_{s.o} \simeq \frac{1}{2M^2} \left( \frac{1}{r} \frac{\partial}{\partial r}
(V^0(r)-S(r))\right) {\bf l} \cdot {\bf s} \; ,
\end{equation}
represents a short-range effect (see also~\cite{chiral1}), determined by the 
condensate structure of QCD. 
The isospin-dependent part of the interaction 
has been studied in the description of the neutron radii. 
Standard relativistic mean-field calculations sistematically
overestimate the difference between neutron and proton radii~\cite{ref1}.
It turns out that the 
density dependence of the isovector channel of the interaction is crucial
in order to reproduce these observables, as also shown in~\cite{DD}.  
\begin{figure}[ht]
    \centering
    \includegraphics*[scale=0.30,angle=0]{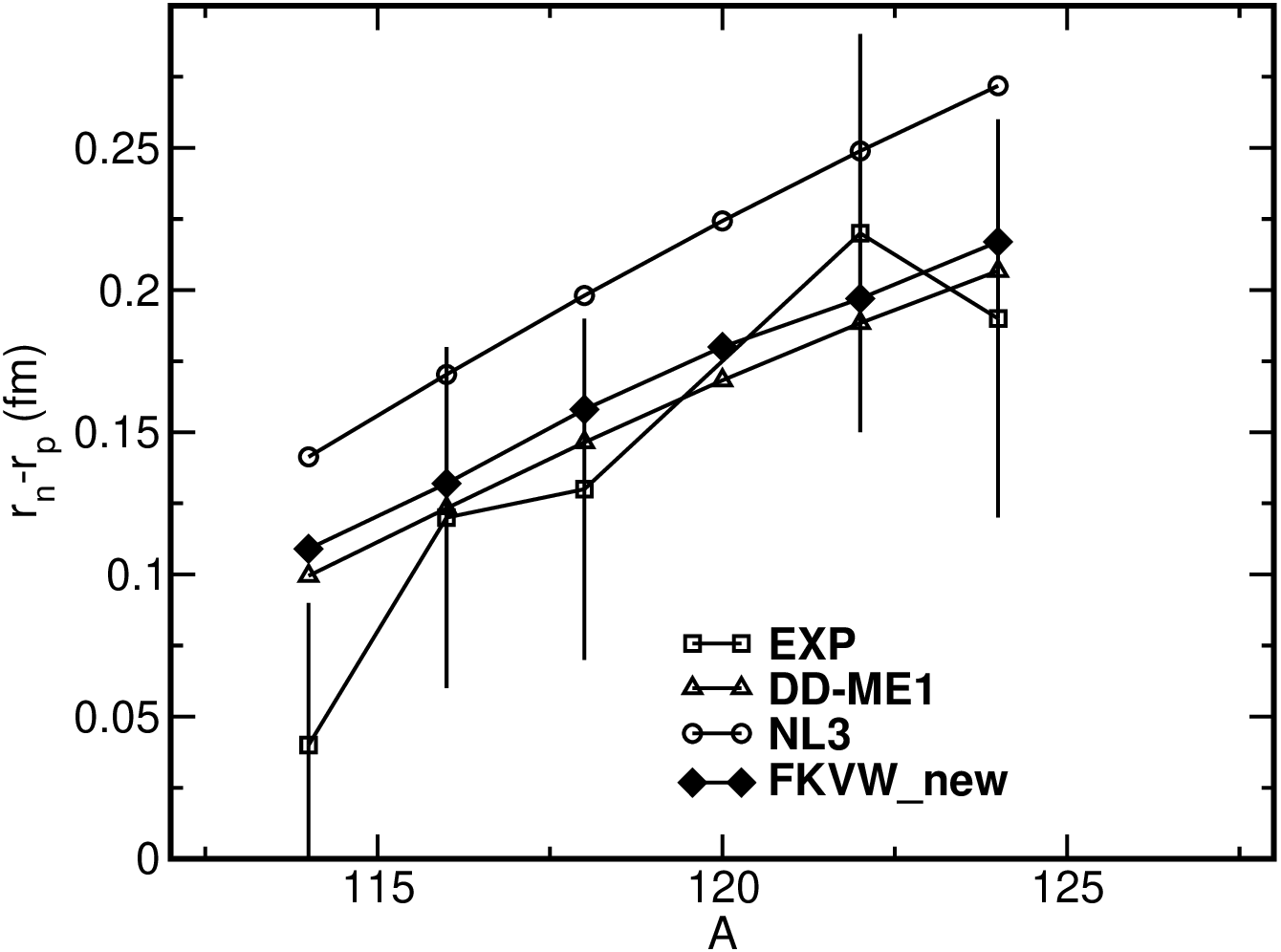}
\caption{Predictions (diamonds) for the differences between 
neutron and proton radii of Sn isotopes, in comparison with NL3 (empty circles), 
DD-ME1 (empty triangles) predictions and experimental data (empty squares)
.\label{fig4}}
\end{figure}
In Fig.~\ref{fig4} we show the calculated values of $r_n-r_p$ for the Sn 
isotope chain in comparison with experimental data~\cite{neutrons}. 
The agreement is excellent, and it is a clear indication that the isovector
channel can be successfully described by pionic fluctuations.

\section{Conclusions}
It has been demonstrated that an approach to nuclear dynamics constrained by
the low-energy sector of QCD provides a quantitative 
description of properties of nuclear matter  
and finite nuclei.


\begin{thebibliography}{0}
\bibitem{ref1} M. Bender, P.-H. Heenen, and P.-G. Reinhard, 
Rev. Mod. Phys. {\bf 75} (2003) 121, P.~Ring,
Prog.\ Part.\ Nucl.\ Phys.\  {\bf 37} (1996) 193, B.~D.~Serot and J.~D.~Walecka,
Int.\ J.\ Mod.\ Phys.\ E {\bf 6} (1997) 515.

\bibitem{nda} J.~J.~Rusnak and R.~J.~Furnstahl,
Nucl.\ Phys.\ A {\bf 627} (1997) 495 .

\bibitem{ref2} T.~D.~Cohen, R.~J.~Furnstahl and D.~K.~Griegel,
Phys.\ Rev.\ Lett.\  {\bf 67} (1991) 961,
R.~J.~Furnstahl, D.~K.~Griegel and T.~D.~Cohen,
Phys.\ Rev.\ C {\bf 46} (1992) 1507,
T.~D.~Cohen, R.~J.~Furnstahl, D.~K.~Griegel and X.~m.~Jin,
Prog.\ Part.\ Nucl.\ Phys.\  {\bf 35} (1995) 221.

\bibitem{ref3} W. Weise, Chiral Dynamics and the Hadronic Phase of QCD, 
Proc. of the International School of Physics  Enrico Fermi  
From Nuclei and their Con- stituents to Stars, Varenna (2002), 
A. Molinari et al., eds., IOS Press, Amsterdam (2003), p. 473-529.

\bibitem{ref4} R.~J.~Furnstahl and B.~D.~Serot,
Comments Nucl.\ Part.\ Phys.\  {\bf 2} (2000) A23 .

\bibitem{ref5} C.~Fuchs, H.~Lenske and H.~H.~Wolter,
Phys.\ Rev.\ C {\bf 52} (1995) 3043, S.~Typel and H.~H.~Wolter,
Nucl.\ Phys.\ A {\bf 656} (1999) 331, T.~Niksic, D.~Vretenar, 
P.~Finelli and P.~Ring, Phys.\ Rev.\ C {\bf 66}, 024306 (2002).

\bibitem{ref7} 
P.~Finelli, N.~Kaiser, D.~Vretenar and W.~Weise,
Eur.\ Phys.\ J.\ A {\bf 17} (2003) 573,
P.~Finelli, N.~Kaiser, D.~Vretenar and W.~Weise,
Nucl.\ Phys.\ A {\bf 735} (2004) 449.

\bibitem{sigma} J.~Gasser, H.~Leutwyler and M.~E.~Sainio,
Phys.\ Lett.\ B {\bf 253} (1991) 252.

\bibitem{quark} A.~Pich and J.~Prades,
Nucl.\ Phys.\ Proc.\ Suppl.\  {\bf 86} (2000) 236,
B.~L.~Ioffe,
Phys.\ Atom.\ Nucl.\  {\bf 66} (2003) 30
[Yad.\ Fiz.\  {\bf 66} (2003) 32].

\bibitem{chiral1}
N.~Kaiser, S.~Fritsch and W.~Weise,
Nucl.\ Phys.\ A {\bf 697} (2002) 255,
S.~Fritsch, N.~Kaiser and W.~Weise,
arXiv:nucl-th/0406038.

\bibitem{chiral2}
N.~Kaiser, S.~Fritsch and W.~Weise,
Nucl.\ Phys.\ A {\bf 700} (2002) 343,
S.~Fritsch, N.~Kaiser and W.~Weise,
Phys.\ Lett.\ B {\bf 545} (2002) 73,
S.~Fritsch and N.~Kaiser,
Eur.\ Phys.\ J.\ A {\bf 17} (2003) 11,
N.~Kaiser, S.~Fritsch and W.~Weise,
Nucl.\ Phys.\ A {\bf 724} (2003) 47.

\bibitem{NL3} G.~A.~Lalazissis, J.~Konig and P.~Ring,
Phys.\ Rev.\ C {\bf 55} (1997) 540 .

\bibitem{DD}T.~Niksic, D.~Vretenar, P.~Finelli and P.~Ring,
Phys.\ Rev.\ C {\bf 66}, 024306 (2002).

\bibitem{neutrons} A. Krasznahorkay {\it et al.}, Phys. Rev. Lett. {\bf 82},
3216 (1999).

\end{thebibliography}
\end{document}